\documentclass[twocolumn,preprintnumbers,amsmath,amssymb,prl]{revtex4}
\usepackage[dvips]{graphicx}
\usepackage{dcolumn}
\usepackage{bm}
\usepackage[dvips]{epsfig}

\begin{document} 

\title{Creep motion of a granular pile induced by thermal cycling}

\author{Thibaut Divoux, Herv\'e Gayvallet and Jean-Christophe G\'eminard.}
\affiliation{Universit\'e de Lyon, Laboratoire de Physique, Ecole Normale Sup\'erieure de
Lyon, CNRS, 46 All\'ee d'Italie, 69364 Lyon cedex 07, France.}

\begin{abstract}
We report a time-resolved study of the dynamics associated with the slow compaction of a granular column submitted to thermal cycles.
The column height displays a complex behavior: 
for a large amplitude of the temperature cycles, the granular column settles continuously,
experiencing a small settling at each cycle;
By contrast, for small-enough amplitude, the column exhibits a discontinuous and intermittent activity:
successive collapses are separated by quiescent periods whose duration is exponentially distributed.
We then discuss potential mechanisms which would account for both the compaction and the transition
at finite amplitude.
\end{abstract}

\maketitle

Literature canonically presents granular materials as a bunch of athermal 
particles \cite{Knight,Philippe,Pouliquen}. Indeed, the energy necessary 
for the grains to hop one over another (a few tenths of millimeter, 
for instance) is roughly 10 orders of magnitude higher than the ambient thermal 
agitation $k_BT$. This is probably one of the reasons why the behavior of a granular 
assembly submitted to temperature fluctuations has received so few attentions. 
Nonetheless, uncontrolled thermal dilations of a granular pile have been reported 
to generate stress fluctuations large enough to hinder reproducible measurements 
of the stress field inside the pile \cite{Vanel,Clement}, and even suspected to 
be the driving factor leading to large-scale "static avalanches" \cite{Claudin}.
Indeed, the slow relaxation and compaction of a granular material, which 
has been hitherto produced by the input of mechanical energy \cite{Richard}, 
can be induced by periodically raising and then lowering the temperature of 
the granular assembly, as recently brought to the fore by Chen and co-workers 
\cite{Chen}. However, the compaction dynamics as well as the basic 
mechanisms at stake remain unknown. In this letter, we address the 
following questions:
What is the dynamic of the top of a granular column submitted to thermal cycling? 
Does this compaction process exhibit features analogous to aging, as 
other compaction processes (tapping, cyclic shear...) do?
And finally, what is the behaviour of the grain assembly in the limit of 
low amplitude temperature-cycles, i.e. well below a cycling amplitude of 40$^{\circ}$C \cite{Chen}?

Compaction of granular materials has been thoroughly studied over the 
past 20 years both experimentally and theoretically \cite{Richard}. It has been achieved by 
direct input of mechanical energy. By far, the most-used device consists of 
a vertical column which is periodically shaken with a controlled vibration 
intensity. Under vibration, a low-density granular-pile compacts towards a 
steady state with a final density that depends on the intensity of the 
vibrations \cite{Knight,Philippe}. Nonetheless, those final states have been 
shown not to depend on the initial conditions (packing fraction) 
and even to be genuine thermodynamic states \cite{Ribiere}. 
In addition to their general interest \cite{Richard}, vibrated granular materials 
have been found to exhibit deep common physical features with out-of-equilibrium 
systems, such as glasses. First, the spectrum of density fluctuations was found 
to be strongly non-Lorentzian \cite{Nowak}, which is a signature of multiple 
timescales in the system. Also, the response of the column to an abrupt
change in shaking intensity is analogous to aging in glasses \cite{Josserand,Kabla}. 
Finally, in the limit of small-intensity vibrations, the relaxation rate becomes 
so slow  that the system cannot reach the steady-state density within the experiment timescale. 
At the microscopic scale, a grain exhibits random motion within confined volumes or cages, 
occasionally a longer exploration corresponding to a change of cage, and rarely jumps, 
during which it moves "significantly" \cite{Ribiere2}.
Cyclic shear deformations of a granular assembly also lead to compaction and 
present the transient cage effects reminding one of colloidal glasses \cite{Nicolas,Pouliquen}. 
Those experimental results have motivated numerous theoretical and numerical works, 
a review of which can be found in \cite{Richard}.\\
\begin{figure}[!h]
\begin{center}
\includegraphics[height=5cm]{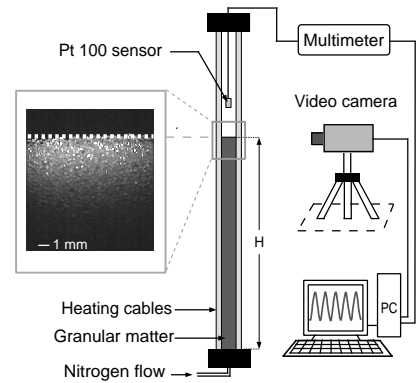}
\end{center}
\caption{\small{{\bf Sketch of the experimental setup.} Inset: picture of the upper part 
of the column. The granular level is indicated by the white dotted-line.}
}
\label{fig.1}
\end{figure}

Up to now, the heating of a granular pile has only been used 
to develop dynamic light-scattering experiments \cite{Djaoui} or 
to probe numerically its impact on the force network \cite{Vargas}.
Here, we report the time-resolved compaction dynamics of a granular column 
submitted to a large number of temperature cycles and shed new 
lights on the key ingredients governing this process.

{\it Experimental set-up -} 
The experimental set-up (Fig.~\ref{fig.1}) consists in a vertical glass tube (height 1.7~m, inner diameter 13~mm)
firmly fastened to a wall into the basement of the physics department, in order to flee
mechanical vibrations. The sample consists of a column (height $H$) of spherical glass beads (diameter $d$)
poured into the tube. An additional gas-input, at the bottom of the column, makes it possible
to loosen the pile thanks to an upward flow of dry nitrogen.
The temperature cycles are imposed by means of a heating cable (Prolabo, 40~W/m) directly taped on
the outer surface of the tube wall. The resulting temperature is measured by means
of a sensor (Pt100, located close to the free surface of the granular material)
and a multimeter (Keithley Intruments, Model 196).  
The free surface of the material, which is illuminated by a red LED (Kingbright, L-793SRC-E,
located inside the tube, above the granular material)
is imaged from the side with a video camera (Panasonic, WV-BP500) connected to
a frame grabber board (Data Translation, DT2255). 
A macro, running under a data-processing software (WaveMetrics, IGOR Pro 4.0), makes it possible
to drive the heating power, to record the resulting variations of the temperature
and to measure accurately the height $H$ from the images: 
A subpixel resolution (namely, less than a tenth of a pixel which typically stands for 5 $\mu $m)
is achieved by considering the average position of the free surface, assumed to correspond to the
inflection point in the vertical intensity-profile averaged over the whole diameter of the tube.
Measurements are performed $20$ times per temperature cycle.

Due to long experimental times, we limit our report to a given diameter
$d = 510 \pm 90~\mu$m of the grains (Matrasur Corp.)
and to a given period $2\pi/\omega = 600$~s of the cycles. 
The cycling period, 10 minutes, is arbitrarily chosen to be small enough to avoid excessively-long
experimental times but large enough to insure that the associated thermal penetration-length
$l_p\equiv \sqrt{ 2\lambda/(\rho\, C \omega)} \simeq 6$~mm is about the tube radius
($\lambda \simeq$~0.2~W m$^{-1}$ K$^{-1}$ and $C \simeq 10^6$~J m$^{-3}$ K$^{-1}$ respectively
denote the thermal conductivity and heat capacity of a typical glass-grains pile \cite{Geminard01}.)
It is here crucial to note that the column is heated homogeneously along its whole length but
that the temperature is likely to vary in the radial direction.

\begin{figure}[h]
\begin{center}
\includegraphics[width=0.9\columnwidth]{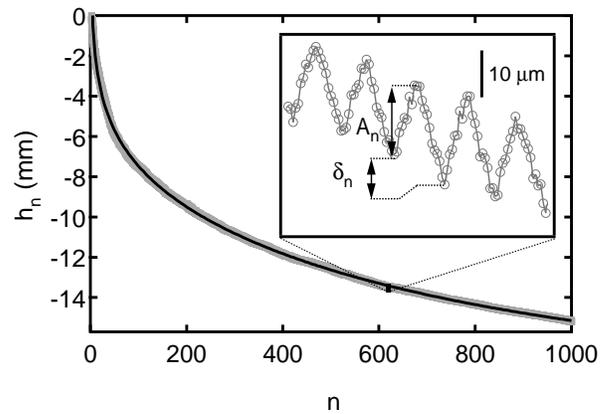}
\end{center}
\caption{\small{\bf Height variation $h_n$ vs. number of cycles $n$.} 
One observes first an exponential behavior at short time followed by a subsequent
logarithmic creep at long time (The black curve corresponds to the test function $h^t_n$ defined in the text.)
Inset: Oscillations of the column height associated with the temperature cycles
($A_n$ and $\delta _n$ are respectively defined to be the amplitude of the increase and the drift of $h_n$
at the cycle $n$.)
For the chosen $\Delta T$, the column settles slightly at each cycle
($H = 140$~cm, $2 \pi/\omega = 600$~s and $\Delta T=10.8^{\circ}$C.) 
} 
\label{fig.2}
\end{figure}

Prior to each experiment, the granular column is prepared in a low-density 
state thanks to the dry-nitrogen upward flow.
The top of the column is then higher than the field imaged by the camera (typically 1~cm above)
and we set the amplitude of the cycles, $\Delta T$, to the largest accessible value, $\Delta T=27.1^{\circ}$ C.
The preparation of the sample ends when the top of the column enters the observation field.
At this point, the granular column is "quenched": The amplitude of the cycles is set to the 
chosen value $\Delta T$ lying bewteen $0$ and $27.1 ^{\circ}$C, which defines the origin of time $t= 0$.
The granular column is subsequently submitted to at least $1000$ cycles (7 days).

{\it First experimental observations -}
Under the action of the temperature cycles the column height decreases:
we report the variation $h_n \equiv H(2 \pi n / \omega) - H(0)$,
where $n$ denotes the time in units of the cycle period or, equivalently when integer, the number
of imposed cycles (Fig.~\ref{fig.2}).
We observe that the thermal-induced compaction is a very slow phenomenon:
after 7 days ($1000$ cycles), the decrease of the height is of about 1.5~cm
(about 1\% of the height $H$), which indicates that the system remains very far
from the maximum compaction
(roughly a decrease of about 10\% of the column height, i.e. $h_n \sim 10$~cm)
within the experimental time.
Accordingly, after the experiment, we checked that a single finger tap produces a 
collapse of the granular height of one centimeter at a rough estimate.
We also checked during 3 days that the height $H$ of the column stays constant when no 
temperature cycles are imposed, which proves that ambient mechanical vibrations
and changes in the room temperature have no (or little) effect in our experimental
conditions.
\begin{figure}[h]
\begin{center}
\includegraphics[width=0.9\columnwidth]{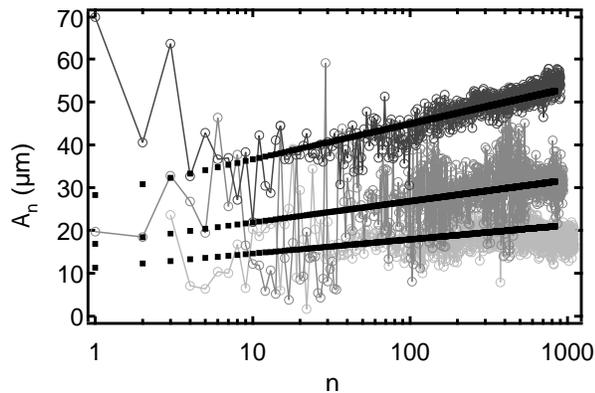}
\end{center}
\caption{\small{\bf Amplitude $A_n$ vs. number of cycles $n$.}
The data are successfully accounted for by
$A_n=\Delta T \, [a_0+b_0$ ln$(n)]$ with $a_0 \simeq 13.8$~nm/K and $b_0 \simeq 2.5$~nm/K 
($H = 140$~cm, $2 \pi/\omega = 600$~s and, from bottom to top,
$\Delta T=10.8, 16.2 {\rm~and~} 27.1^{\circ}$C.)}
\label{fig.dilation}
\end{figure}

Before we discuss the compaction dynamics, it is relevant to determine
which part of the column is involved in the phenomenon.
Our measurements are accurate enough for observing oscillations of $H$
associated with the temperature variations (Fig.~\ref{fig.2}, inset).
We observe on the raw data that the amplitude $A_n$, which is proportional to 
$\Delta T$, increases logarithmically with $n$ (Fig.~\ref{fig.dilation}).
The oscillations of $H$ are due to the thermal dilation of both the tube
and the granular material. In order to assess the contribution of the granular
material, we first determine the amplitude,
$\delta h_t(z)$, of the tube displacement in the laboratory frame as a function
of the height $z$ (origin at the bottom of the column) by marking its outer wall.
The amplitude $\delta h_t(z)$ is found to be linear in $z$ and
the slope provides us with an estimate of the linear thermal-expansion coefficient, 
$\kappa = (3.6 \pm 0.4)\,\times 10^{-6}$~K$^{-1}$, of the tube material.
Then, considering the relative variation of the inner volume,
we write the relation  between the amplitude $A_n$ and
the relative variation, $\delta V_g/V_g$, of the volume $V_g$
of the granular material:
$A_n - \delta h_t(0) = H (\delta V_g/V_g - 2 \kappa \Delta T)$.
Experiments performed for different height $H$ 
demonstrate that $A_n - \delta h_t(0)$ is proportional to $H$,
which shows that $\delta V_g/V_g$ is independent of $H$.
We thus infer that {\it the whole height $H$ of the granular column is
involved in the observed oscillations of the free surface}.

\begin{figure}[t]
\begin{center}
\includegraphics[width=0.9\columnwidth]{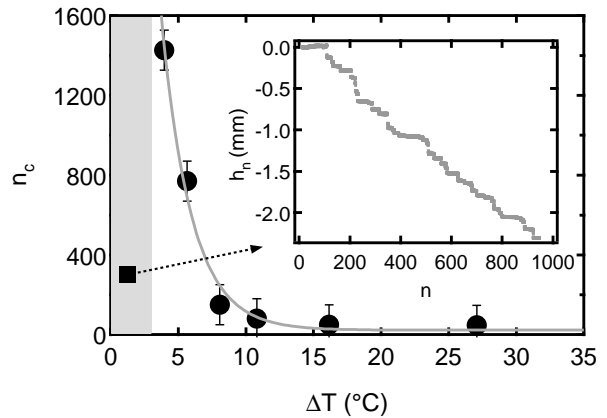}
\end{center}
\caption{\small{\bf Characteristic number $n_c$ vs. amplitude $\Delta T$.}
The characteristic number of cycles $n_c$ increases drastically when $\Delta T$ is 
decreased and even seems to diverge for $\Delta T_c~\simeq~3^{\circ}$C. 
Inset: For $\Delta T < \Delta T_c$, the column settles, by jumps, linearly with time.
For readibility, we only display data obtained during the first half of the experiment
duration (14 days, $H = 140$~cm, $2 \pi/\omega = 600$~s and $\Delta T=2.8^{\circ}$C.)}
\label{fig.3}
\end{figure}

{\it Compaction dynamics -}
For a large $\Delta T$ (typically more than 3$^{\circ}$C), the column systematically
compacts during the decrease of the temperature, at each cycle, and the amplitude $\delta_n$
of the collapses (defined in Fig.~\ref{fig.2}, inset) decreases when $n$ increases,
which accounts for the overall behavior of $H$ as a function of $n$.
We show here that the results obtained by Chen and co-workers for $\Delta T > 40^\circ$C \cite{Chen}
hold true at much smaller $\Delta T$. 
Besides, we resolve the column dynamics and, this, over a significant longer duration.
Here, the column height first decreases exponentially to reach,
after the characteristic number $n_c$ of cycles, a logarithmic behavior at long time (Fig.~\ref{fig.2}).
This response to the thermal quenching is very similar to the one the system exhibits to step strain
solicitation \cite{Brujic}. 
We estimated $n_c$ from the experimental data by
interpolating $h_n$ with the test function $h_n^t\equiv\,h_0+h_e\,\exp{(-n/n_c)} + h_l\,\ln(n)$.
In addition, we checked that the shortest characteristic time obtained from the interpolation
by a double exponential \cite{Barker} is consistent with $n_c$ and presents the same behavior as a function of $\Delta T$.
We observe that $n_c$ drastically increases when $\Delta T$ is decreased.
Moreover, $n_c$ even seems to diverge at $\Delta T_c$
(We get $2.8 < \Delta T_c < 4.0\,^{\circ}$C, Fig.~\ref{fig.3}), which prefigures a
qualitatively-different dynamics of the column compaction at a lower amplitude.

\begin{figure}[h]
\begin{center}
\includegraphics[width=0.9\columnwidth]{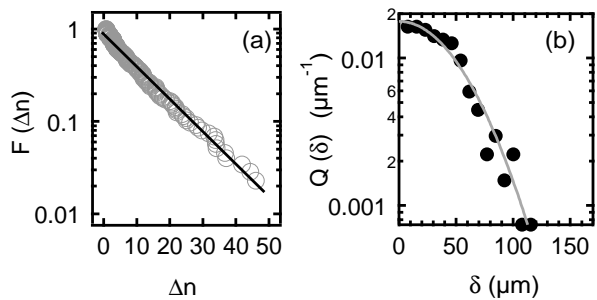}
\end{center}
\caption{\small{\bf Statistics below the transition}.
(a)~:~The cumulated probability $F(\Delta n) \simeq \exp{(-\Delta n/\Delta n^*)}$ is an exponential function 
of $\Delta n$ and the characteristic number $\Delta n^* \simeq 12$.
(b)~:~The probability distribution $Q(\delta) \simeq \exp{[-(\delta/\delta^*)^2]}$ is a Gaussian function whose
width $\delta^* = 63\pm 4\,\mu$m.
The statistical data reported herein correspond to the 175 collapses observed within the experimental time
($14$ days, $H = 140$~cm, $2 \pi/\omega = 600$~s and $\Delta T={2.8}^{\circ}$C).}
\label{fig.4}
\end{figure}

For $\Delta T < \Delta T_c$, one indeed observes that
the column is not flowing regularly anymore,
but evolves by successive collapses separated by rest periods (Fig.~\ref{fig.3}, inset).
The overall compaction velocity is constant (steady regime) and it is therefore relevant
to consider the probability distribution $P(\Delta n)$ of the number of cycles $\Delta n$
between two successive collapses (Fig.~\ref{fig.4}).
However, due to the small number of collapses, $N = 175$, within the experimental time (14 days),
the behavior of $P(\Delta n)$ is difficult to assess directly and
we consider the cumulated probability, $F(\Delta n) \equiv \int_{\Delta n}^{\infty} P(x)$~d$x$, instead \cite{Remark1}.
We obtain $F(\Delta n) \simeq \exp{(-\Delta n/\Delta n^*)}$ with $\Delta n^* \simeq 12$ for
$\Delta T={2.8}^{\circ}$C (Fig.~\ref{fig.4}a).
The delay $\Delta n$ between two successive collapses is thus observed to be
exponentially distributed. By contrast, the amplitude of the corresponding collapses, $\delta_n$,
exhibits a Gaussian probability-distribution $Q(\delta)$ whose width is estimated to be
$\delta^* = 63 \pm 4\,\mu$m (Fig.~\ref{fig.4}b), roughly a tenth of the grain diameter.
The compaction of the material is thus shown to result from collapses that are randomly distributed in time.

{\it Discussion and conclusion -} 
The mechanisms leading to the compaction could be either extrinsic (the dilation of the container at stake)
or intrinsic (only due to the dilation of the grains.)
Chen {\it et al.} suggested that the compaction could be due to
"the difference between the thermal expansion of the container and of the grains" \cite{Chen},
thus to an extrinsic mechanism. 
In our case, the beads ($\kappa_b = 9 \times 10^{-6}$ K$^{-1}$) and the container 
($\kappa = 3.6 \times 10^{-6}$ K$^{-1}$) are made of glass and present roughly 
the same thermal expansion coefficient \cite{Remark2}. Thus, we do not 
think that the thermal dilation of the container is the primary cause of the compaction.
This conclusion is supported by preliminary results obtained by heating periodically
the granular column along its axis \cite{geminard03} and the recent simulations by Vargas 
and McCarthy \cite{Vargas}. 
Dealing with an intrinsic mechanism, we guess that an overall temperature
change would only produce a homogeneous dilation of the pile and no spatial reorganisation
of the grains. We thus propose the following scenario, compatible with previous results \cite{Chen}:
due to the radial thermal-gradient of amplitude $\Delta T/l_p$, the grains are subjected to
a periodic shear, $\gamma$, along the vertical which induces local rearrangments and, accidentally,
internal avalanches.
This mechanism is supported by the estimate of the shear $\gamma = \kappa_b (\Delta T/l_p) L$,
where $L$ is a characteristic size which ranges from the grain- to the tube-diameter.
We obtain, for $L \in [0.510,13]$~mm, that the relative displacement of two neighbor grains 
$\Gamma \equiv \gamma d/\Delta T \in [0.4,10]$~nm/$^\circ$C whereas,
using an AFM, we estimated the maximum size of the asperities associated with the surface roughness
of the beads, $s \simeq 100$~nm
(Invoking the surface roughness is supported by numerical simulations:
for perfectly smooth grains the maximum compacity is reached after a few cycles \cite{Vargas},
which contrasts with our observations.) 
In this framework, the continuous compaction would thus correspond to amplitudes $\Delta T$
such as $\Gamma \Delta T > s$, whereas jumps would be observed for $\Delta T < \Delta T_c \simeq s/\Gamma$.
From the experimental $\Delta T_c \simeq 3^\circ$C, we could infer that the typical length at stake, $L$,
is about the tube radius and, thus, the transition a finite-size effect.
Due to long experimental times, we limited our report to results
which clearly demonstrate the efficiency of the techniques.
The dependance of the critical amplitude $\Delta T_c$ on the column
height and diameter as well as on the cycling frequency and grain roughness
is necessary to corroborate or discard the proposed mechanism.
In particular, the sensitivity to the frequency $\omega$
will indicate the role played by the temperature gradient.
We also plan to locate the origin of the collapses by listening to the
associated noise and to observe the local displacement by optical means.
This way, we hope to correlate the macroscopic behavior 
to the local dynamics as already done for tapping \cite{Ribiere2,Lumay}
and cycling shear experiments \cite{Pouliquen,Marty}.

{\bf Acknowledgments} We thank \'Eric Bertin, Sergio Ciliberto,
and Jorge Kurchan for fruitful discussions.


\begin{thebibliography}{}
\bibitem{Remark2} In \cite{Chen}, $\kappa /\kappa_b$ and $\kappa_b/\kappa$ are systematically larger than 13.
\end{thebibliography}
\end{document}